\documentclass[10pt,english,showpacs,letterpaper,prl,amsfonts,amssymb,oneside,balancelastpage,notitlepage,twocolumn]{revtex4}
\usepackage{fontenc}
\usepackage[latin1]{inputenc}
\usepackage{amsmath}
\usepackage{graphicx}
\usepackage{amssymb}
\usepackage{amsfonts}
\usepackage{babel}

\usepackage{epsfig}
\usepackage{epstopdf}

\providecommand{\U}[1]{\protect\rule{.1in}{.1in}}

\begin{document}
\title{$\mathrm{N}$-dimensional alternate coined quantum walks from a dispersion
relation perspective}
\author{Eugenio Rold\'an$^{(1)}$, Carlo Di Franco$^{(2,3)}$, Fernando Silva$^{(1)}$, and
Germ\'an J. de Valc\'arcel$^{(1)}$, }
\affiliation{$^{(1)}$Departament d'\`Optica, Universitat de Val\`encia, Dr. Moliner 50,
46100-Burjassot, Spain}
\affiliation{$^{(2)}$Departamento de Qu\'imica F\'isica, Universidad del Pa\'is Vasco-Euskal
Herriko Unibertsitatea, Apartado 644, E-48080 Bilbao, Spain}
\affiliation{$^{(3)}$CTAMOP, School of Mathematics and Physics, Queen's University, Belfast BT7 1NN, United Kingdom}

\begin{abstract}
We suggest an alternative definition of $\mathrm{N}$-dimensional coined
quantum walk by generalizing a recent proposal [Di Franco \textit{et al.},
Phys. Rev. Lett. \textbf{106}, 080502 (2011)]. This $\mathrm{N}$-dimensional
alternate quantum walk, $\mathrm{AQW}_{(\mathrm{N})}$, in contrast with the
standard definition of the $\mathrm{N}$-dimensional quantum walk,
$\mathrm{QW}_{(\mathrm{N})}$, requires only a coin qubit. We discuss the
quantum diffusion properties of $\mathrm{AQW}_{(2)}$ and $\mathrm{AQW}_{(3)}$
by analyzing their dispersion relations that reveal, in particular, the
existence of diabolical points. This allows us to highlight interesting
similarities with other well-known physical phenomena. We also demonstrate
that $\mathrm{AQW}_{(3)}$ generates considerable genuine multipartite entanglement. Finally,
we discuss the implementability of $\mathrm{AQW}_{\left(  \mathrm{N}\right)}$.

\end{abstract}

\maketitle

In both its standard forms, the coined \cite{discreteQW} and the continuous
one \cite{continuousQW}, quantum walk is the quantum version of a
classical random process, described by the diffusion and the telegrapher's
equations, respectively \cite{Strauch}. In the coined quantum walk -- the
process we consider here -- there is a system (the walker) that undergoes a
conditional displacement, to the right or the left, depending on the output of a
\textit{coin throw}, as in the random walk. But differently from its classical
counterpart, here both coin and walker are quantum in nature.
The one-dimensional coined
quantum walk -- $\mathrm{QW}_{\mathrm{(1)}}$ for short -- has been studied
from many different perspectives, especially from the quantum computational
point of view \cite{reviews}. In the last few years, quantum walks have also received
increasing experimental attention \cite{Bouwmeester,experiments,Schreiber10},
including cases with more than one particle \cite{multiparticle}.

The situation is quite different when dealing with $\mathrm{N}$-dimensional
quantum walks, $\mathrm{QW}_{\mathrm{(N)}}$ for short. They were first
discussed by Mackay \textit{et al}., who introduced them in complete analogy
with $\mathrm{QW}_{\left(  1\right)  }$ \cite{Mackay02} (see also Ref.
\cite{Moore}). As defined in Ref. \cite{Mackay02}, $\mathrm{QW}_{\mathrm{(N)}
}$ requires the use of a $2^\mathrm{N}$-dimensional qudit as coin, as well as a
coin operator represented by a $\mathrm{2^N\times2^N}$ unitary matrix. This
introduces increasing complexity in the process as $\mathrm{N}$ grows,
especially from the experimental viewpoint \cite{Soriano,Schreiber12}, but
also from the theoretical one \cite{2DQW,Inui}. However, Di Franco \textit{et
al. }\cite{DiFranco11a} have recently proposed an alternative two-dimensional
quantum walk, namely the \textit{alternate} quantum walk -- $\mathrm{AQW}$ for
short -- that is simpler than the standard one. In $\mathrm{AQW}$
the coin is a single qubit, as in $\mathrm{QW}_{\mathrm{(1)}}$, and each time
step is divided into two halves: in the first one the coin is thrown (i.e., a
Hadamard transformation is applied on the coin qubit) and the conditional
displacement along
$x$ is performed; then, in the second
half of the time step, the coin is thrown again and the conditional
displacement along
$y$ is performed. Hence, in $\mathrm{AQW}$, the
four-dimensional qudit of $\mathrm{QW}_{\mathrm{(2)}}$ is replaced by a single
qubit, the price paid for that being to double the number of sub-steps per
single time step. Quite unexpectedly, $\mathrm{AQW}$ reproduces the same spatial
probability distributions of $\mathrm{QW}_{\left(  2\right)  }$ when the
Grover coin is used -- Grover-$\mathrm{QW}_{\mathrm{(2)}}$ for short -- for a
set of particular initial conditions, precisely those for which the
characteristic localization of Grover-$\mathrm{QW}_{\mathrm{(2)}}$ does not
occur. In Refs. \cite{DiFranco11a,DiFranco11b} analytical demonstrations of
the (partial) equivalence between Grover-$\mathrm{QW}_{\left(  2\right)  }$
and \textrm{AQW} are given.

Here we generalize \textrm{AQW} to $\mathrm{N}$-dimensions: we define the
$\mathrm{N}$-dimensional alternate coined quantum walk -- $\mathrm{AQW}
_{\mathrm{(N)}}$ for short -- as a quantum walk in which the time steps are
divided into $\mathrm{N}$ sub-steps. In each of these sub-steps, the coin
throw is followed by the conditional displacement along one of the
$\mathrm{N}$ dimensions. In a single time step of \textrm{AQW}$_{\left(
\mathrm{N}\right)  }$, the qubit-coin is therefore thrown $\mathrm{N}$ times,
but this is clearly simpler than throwing a single $\mathrm{2^N}$-dimensional
coin, from both the experimental and analytical points of view, in particular
for large $\mathrm{N}$ (a more detailed discussion about the possible scaling
of errors strongly depends on the physical setting exploited for the
realization of the scheme). We show below that, besides a simpler experimental
implementability as compared with $\mathrm{QW}_{\left(  \mathrm{N}\right)  }$,
\textrm{AQW}$_{\left(  \mathrm{N}\right)  }$ has a rich dynamics that is also
easy to understand. We provide
some analytical results concerning the evolution of the probability
distribution for cases $\mathrm{N}=2,3$, paying special attention to the
processes' dispersion relations. These reveal the existence of diabolical
points (DPs in the following), conical intersections involving a degeneracy
\cite{BerryWilkinson}, that allow us to highlight interesting similarities with other
well-known physical phenomena. As a striking example, we present the
homogeneous propagation of an initially extended state with perfect circular
symmetry, in relation to the ring shape of the beam exiting a biaxial crystal under
conical refraction conditions \cite{Berry}. Further physical contexts where DPs
appear are given in the remainder of the paper. 
{\it An analysis from this viewpoint has not been
performed so far and is useful to broaden our knowledge of the particular
scheme under investigation}. We also demonstrate the generation of considerable genuine multipartite entanglement in
\textrm{AQW}$_{\left(  3\right)  }$. We conclude this paper with a brief
discussion concerning the experimental implementability of \textrm{AQW}
$_{\left(  \mathrm{N}\right)  }$.

\section{The model}

In order to introduce \textrm{AQW}$_{\left(  \mathrm{N}\right)  }$ formally,
let $\left\vert \psi\right\rangle _{t}\ $ represent the state of the system at
(discrete) time $t$. The vector $\left\vert \psi\right\rangle _{t}\ $is defined in
the compound Hilbert space $\mathcal{H}_{P}\otimes\mathcal{H}_{C}$, where
$\mathcal{H}_{P}$ and $\mathcal{H}_{C}$ are the Hilbert spaces for the
lattice sites and coin qubit, respectively. With $\mathcal{H}_{P}$ and
$\mathcal{H}_{C}$ spanned
by $\left\{  \left\vert \vec{x}\right\rangle
,\text{ }\vec{x}\in
\mathbb{Z}
^{N}\right\}  $ and $\left\{  \left\vert
c\right\rangle ,c=u,d\right\}  $, we can write
$\left\vert \psi\right\rangle _{t}=\sum_{\vec{x}}\sum_{c=u,d}c_{\vec{x}
,t}\left\vert \vec{x};c\right\rangle$ 
with $\left\vert \vec{x};c\right\rangle =\left\vert \vec{x}\right\rangle
\otimes\left\vert c\right\rangle $, where $c_{\vec{x},t}$ is the probability
amplitude for the walker to be at site $\vec{x}=\left(  x_{1},\ldots
,x_{N}\right)  $ at time $t$ with the coin in state $c$. The probability of
finding the walker at site $\vec{x}$ at time $t$ is $P_{\vec{x},t}=\left\vert
u_{\vec{x},t}\right\vert ^{2}+\left\vert d_{\vec{x},t}\right\vert ^{2}$.

The state evolves as $\left\vert \psi\right\rangle _{t+1}=\hat{U}_{\left(
N\right)  }\left\vert \psi\right\rangle _{t}$, with $\hat{U}_{\left(
N\right)  }=\hat{D}_{N}\hat{C}_{N}\hat{D}_{N-1}\hat{C}_{N-1}\ldots\hat{D}
_{1}\hat{C}_{1}$ a unitary operator. Here, operator $\hat{C}_{i}$ is the
\textit{coin operator}, acting only in $\mathcal{H}_{C}$, whose more general
form is
$\hat{C}_{i}=\cos\theta_{i}\left(\left\vert u\right\rangle\left\langle
u\right\vert -e^{i\left(  \alpha_{i}+\beta_{i}\right)  }\left\vert
d\right\rangle \left\langle d\right\vert \right) +\sin\theta_{i}\left(  e^{i\alpha_{i}}\left\vert u\right\rangle
\left\langle d\right\vert +e^{i\beta_{i}}\left\vert d\right\rangle
\left\langle u\right\vert \right)$ 
with $\left(  \alpha_{i},\beta_{i},\theta_{i}\right)  $ arbitrary reals and
$i=1,\ldots,N$ (notice that, for $\alpha_{i}=\beta_{i}=0$ and $\theta_{i}
=\pi/4$, $\hat{C}_{i}$ is just the Hadamard transformation). $\hat
{D}_{i}$ is the conditional displacement operator along direction $x_{i}$,
which we write as
$\hat{D}_{i}=\sum_{\vec{x}\in
\mathbb{Z}
^{N}}\left[  \left\vert \vec{x}+\vec{n}_{i};u\right\rangle \left\langle
\vec{x};u\right\vert +\left\vert \vec{x}-\vec{n}_{i};d\right\rangle
\left\langle \vec{x};d\right\vert \right]$,
where $\vec{n}_{i}$ is the unit vector along direction $x_{i}$.

Equation $\left\vert \psi\right\rangle _{t+1}=\hat{U}_{\left(  N\right)
}\left\vert \psi\right\rangle _{t}$ can be expressed as a
map relating probability amplitudes $c_{\vec{x},t+1}$ with $\tilde{c}_{\vec
{x}^{\prime},t}$, where $c,\tilde{c}=u,d$ and $\vec{x}^{\prime}$ are nearest
neighbors of $\vec{x}$. This map admits two plane-wave solutions of the form
$\operatorname{col}\left(  u_{\vec{x},t},d_{\vec{x},t}\right)  _{\pm}
=\vec{\varphi}_{\vec{q},\pm}\exp\left[  i\left(  \vec{q}\cdot\vec{x}
-\omega^{\left(  \pm\right)  }t\right)  \right]  $, where $\vec{\varphi}
_{\vec{q},\pm}=\operatorname{col}\left(  u_{\vec{q},\pm},d_{\vec{q},\pm
}\right)  $\ are time independent vectors, $\vec{q}$ is the pseudo-momentum
with $q_{i}\in\left(  -\pi,\pi\right]  $, and $\omega^{\left(  \pm\right)  }$
are two frequencies determined by the dispersion relation.
The dispersion relation is most relevant because $\left\vert \psi\right\rangle
_{t}$ is entirely determined by it, given $\left\vert \psi\right\rangle
_{t=0}$. Moreover, when the initial state extends over a finite set of points
in the lattice, especially when it is modulated by a \textit{smooth} function
of space, the dispersion relation is particularly useful for predicting the
evolution of the initial wave packet, due to the relatively well-defined group
velocity (given by the local gradient of the dispersion relation curve). In
this case, long-wavelength continuous approximations are very well suited and
useful for envisaging the long time behavior of the probability distribution.
This has been discussed in detail for $\mathrm{QW}_{\left(  1\right)  }$ in
Ref. \cite{deValcarcel10} (see also Refs. \cite{Strauch,Knight}). Here, we
will limit ourselves to a qualitative discussion of what the dispersion
relation suggests for cases $\mathrm{N}=2,3$.

\section{Dispersion relation for two-dimensional alternate quantum walk}

\begin{figure}[t]
\centerline{\psfig{figure=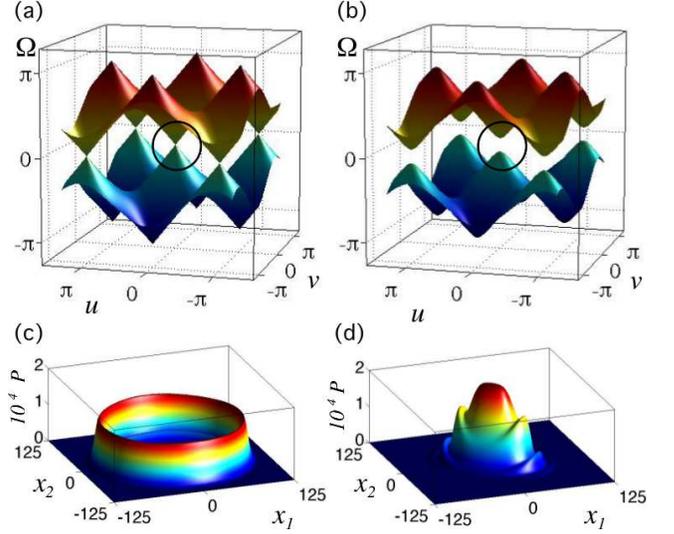,width=8.5cm}}
\caption{(Color online) 3D views of the two branches of Eq. (\ref{disp(2)}), for
$\theta_{1}=\theta_{2}=\pi/4$ (a) and $\theta_{1}=\pi/4\ne\theta_{2}
=\pi/3$ (b). Propagation in $\mathrm{AQW}_{(2)}$ after $90$ time steps of an
initial state with a Gaussian probability distribution of width $\sigma
_{HWHM}=7$ and coin state $\operatorname{col}(1/\sqrt{2},i/\sqrt{2})$
equal for all populated sites, for $\theta_{1}=\theta_{2}=\pi/4$ (c)\ and
$\theta_{1}=\pi/4\ne\theta_{2}=\pi/3$ (d).}
\label{fig1}
\end{figure}

Let us first consider \textrm{AQW}$_{\mathrm{(2)}}$. By proceeding as stated
above, one obtains the following dispersion relation:
\begin{equation}
\cos\Omega=c_{1}c_{2}\cos\left(  u+v\right)  +s_{1}s_{2}\cos\left(
u-v\right)  , \label{disp(2)}
\end{equation}
where $c_{i}=\cos\theta_{i}$, $s_{i}=\sin\theta_{i}$, $u=q_{1}+\left(
\beta_{1}+\alpha_{2}\right)  /2$, $v=q_{2}+\beta_{2}/2$, and $\Omega
=\omega-\left(  \beta_{1}+\beta_{2}+\alpha_{2}\right)  /2$. Notice that phase
$\alpha_{1}$ does not appear in Eq. (\ref{disp(2)}); hence it is irrelevant.
As for the phases that do appear, they just entail a translation of the
frequency and spatial quasi-momentum. Only $\theta_{1}$ and $\theta_{2}$ are
dynamically relevant parameters.
Figure \ref{fig1}(a) presents the dispersion curves for $\theta_{1}
=\theta_{2}=\pi/4$. The most relevant features are the existence of a number
of saddle-points (for which the group velocity is zero), together with regions
of maximum slope (which equals 0.5, hence the maximum velocity in
\textrm{AQW}$_{\left(  2\right)  }$) and, most importantly, the existence of
DPs. Interestingly, when $\theta_{1}\neq\theta_{2}$, the DPs disappear, as shown in Fig. \ref{fig1}(b).
This suggests that the existence of DPs could be particularly sensitive to
decoherence effects in the coin mechanism.

The dynamics around the points of null and maximum slope can be analyzed as in
Ref. \cite{deValcarcel10} for $\mathrm{QW}_{\left(  1\right)  }$, i.e., in
terms of known solutions of simple linear wave equations. One can then
envisage more or less straightforward generalizations of the results there
discussed to the two-dimensional case (we study this elsewhere for
Grover-\textrm{QW}$_{\left(  2\right)  }$ \cite{Marga}). But the existence of
DPs is particularly appealing and constitutes a qualitative difference with
the one-dimensional case. This geometric object, the DP (that takes its name
from the diabolo-like shape of the conical intersection), appears in physics
in quite different contexts such as, for instance, quantum triangular billiards
\cite{BerryWilkinson}, conical refraction in crystal optics \cite{Berry}, the electronic spectrum of
polyatomic molecules \cite{spectra}, or
the dispersion relations for massless fermions (Dirac electrons) in QED and
for electrons in graphene \cite{Bostwick,CastroNeto} or optical lattices \cite{optical lattices}. The diabolo is
associated with some remarkable phenomena appearing in those systems.
{\it As for a given initial condition the
dispersion relation determines the evolution of the system (in the absence of
dissipation), the existence of DPs establishes a strong link between the
evolution properties in these contexts}. Let us remark the fact
that the quantum walk, differently from the continuous systems mentioned above, is defined in a discrete Hilbert space.

For what we have stated so far, we can expect to find in the dynamics of \textrm{AQW}$_{\left(  2\right)  }$
some parallelism with phenomena present in the aforementioned systems. In this
sense, Fig. \ref{fig1}(c) shows the propagation of an initially extended state [with a
Gaussian probability distribution of $\sigma_{HWHM}=7$ and coin state $\operatorname{col}\left(  1,i\right)  /\sqrt{2}$
equal for all populated sites] after $90$ time steps. A homogeneous ballistic propagation from
the origin with perfect circular symmetry is clearly visible, which strongly recalls
the ring shape of the beam exiting a biaxial crystal under conical refraction
conditions \cite{Berry}. Indeed a careful analysis reveals that the fine structure of the
probability distribution in Fig. 1(c) is very similar to that appearing in
conical refraction (the so-called Pogendorf rings \cite{Berry}), a result to
be reported elsewhere with full mathematical details \cite{Marga}.
It is interesting to compare this dynamics with that
shown in Fig. \ref{fig1}(d), which has been obtained for the same initial conditions
but with $\theta_{1}=\pi/4\neq\theta_{2}=\pi/3$. In this case the diabolo is
lost and the branches of the dispersion relation show a parabolic shape [see
Fig. \ref{fig1}(b)] that leads to evolutions typical of linear optical diffraction
\cite{deValcarcel10}. We want to mention that a similar controlled
disappearance of the DP has been experimentally observed in graphene
\cite{Bostwick}.

Before moving to a higher dimension, let us revisit the relation between
\textrm{AQW}$_{\mathrm{(2)}}$ and Grover-$\mathrm{QW}_{\mathrm{(2)}}$ from the
dispersion relation perspective. The dispersion relation for
Grover-$\mathrm{QW}_{\mathrm{(2)}}$ consists of four sheets, because the coin
space is four-dimensional, and can be found in Refs. \cite{Inui,Marga}. In our
notation, they read $\omega_{1,2}=0,\pi$ and $\omega_{3,4}=\pm\arccos\left[
\left(  \cos u+\cos v\right)  /2\right]  $. Remarkably, the two sheets
$\omega_{3,4}$ coincide with those of \textrm{AQW}$_{\left(  2\right)  }$ [Eq.
(\ref{disp(2)})] for $\theta_{1}=\theta_{2}=\pi/4$ after identifying $\left(
u,v\right)  $ in Grover-$\mathrm{QW}_{\left(  2\right)  }$ with $\left(
u+v,u-v\right)  $ in \textrm{AQW}$_{\left(  2\right)  }$, i.e., the two
dispersion relations coincide for these parameters up to a $\pi/4$ rotation of
the pseudo-momentum. The two other roots in Grover-$\mathrm{QW}_{\mathrm{(2)}
}$, $\omega_{1,2}=0,\pi$, are constant, which means that the projections of
the initial state onto the corresponding eigenvectors will not evolve in time.
This is the origin of localization in Grover-$\mathrm{QW}_{\left(  2\right)
}$ for most initial coin states, as already noticed in Ref. \cite{Inui}. We
conclude that, whenever the initial state in Grover-$\mathrm{QW}_{\left(
2\right)  }$ does not project onto the eigenvectors governed by $\omega_{1,2}
$, Grover-$\mathrm{QW}_{\left(  2\right)  }$ and \textrm{AQW}$_{\left(
2\right)  }$ are isomorphous for $\theta_{1}=\theta_{2}=\pi/4$. This is our
proof of the (partial) equivalence between the two versions of the process.

\section{Dispersion relation for three-dimensional alternate quantum walk}

Let us now move to \textrm{AQW}$_{\left(  3\right)  }$. The dispersion
relation is governed by
\begin{align}
\sin\Omega &  =c_{1}\left[  c_{2}c_{3}\sin\left(  u\!+\!v\!+\!w\right)  +s_{2}
s_{3}\sin\left(  u\!-\!v\!+\!w\right)  \right]  \nonumber\\
&  +s_{1}\left[  c_{2}s_{3}\sin\left(  u\!+\!v\!-\!w\right)  -s_{2}c_{3}\sin\left(
u\!-\!v\!-\!w\right)  \right]  ,
\label{disp3}
\end{align}
with $c_{i}\!=\!\cos\theta_{i}$, $s_{i}\!=\!\sin\theta_{i}$, and $\left(
u,v,w,\Omega\right)  \!=\!\left(  q_{1},q_{2},q_{3},\omega\right)  +\left(  \delta
q_{1},\delta q_{2},\delta q_{3},\delta\omega\right)  $. Here $2\delta
q_{1}=\alpha_{1}+\beta_{2}$, $2\delta q_{2}=\alpha_{2}+\beta_{3}$, $2\delta
q_{3}=\alpha_{3}+\beta_{1}$, and $\delta\omega=-\left(  \delta q_{1}+\delta
q_{2}+\delta q_{3}\right)  .$ From Eq. (\ref{disp3}) two dispersion relations are
obtained, namely $\omega^{\left(  +\right)  }=\Omega$ and $\omega^{\left(
-\right)  }=\pi-\omega^{\left(  +\right)  }$. As in \textrm{AQW}$_{\left(
2\right)  }$, some phases in $\hat{C}_{i}$ are absent in the dispersion
relation (hence they are irrelevant) and the effect of the rest of phases in
$\hat{C}_{i}$ is just a displacement of the dispersion relation surfaces in
the $\left\{  q_{1},q_{2},q_{3},\omega\right\}  $ space. Equation (\ref{disp3}) is
simpler for $\theta_{i}=\pi/4$, $i=1,2,3$.
In this case there are eight degeneracies (occurring when $\omega^{\left(
+\right)  }=\omega^{\left(  -\right)  }=\pi/2$) at $\left(  u_{DP}
,v_{DP},w_{DP}\right)  =\{\left(  a,a,a\right)\!  ,\!\left(  -a,-a,b\right)
\!,\!\left(  -a,b,-a\right)\!  ,\!\left(  b,-a,-a\right)  \}$, with $a,b=\pi/4,3\pi/4$
and $a\neq b$. These are the three-dimensional (3D) equivalents of the DPs discussed for
\textrm{AQW}$_{\mathrm{(2)}}$. As in \textrm{AQW}$_{\mathrm{(2)}}$, the DPs
disappear when phases $\theta_{i}$ are different. We mention that several
crystallographic structures have recently been proposed for obtaining 3D DPs
(Dirac-semimetal in 3D) \cite{Young12}.

\begin{figure}[t]
\centerline{\psfig{figure=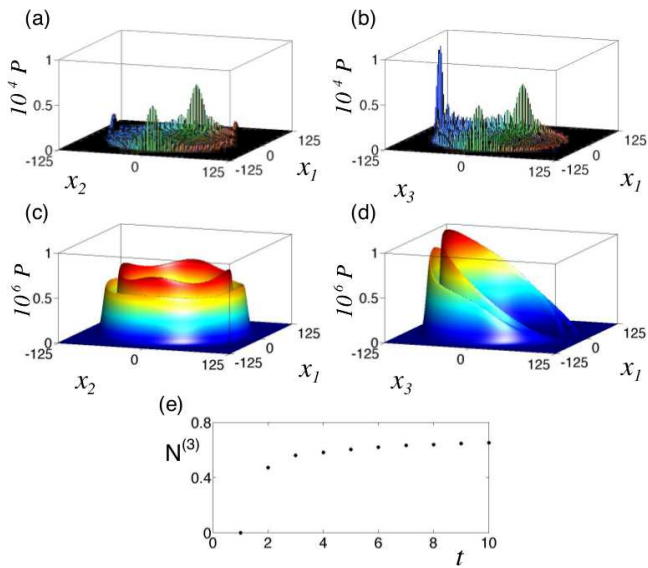,width=8.5cm}}
\caption{(Color online) Propagation in $\mathrm{AQW}_{(3)}$ after $90$ time steps of
(a,b) a localized initial state, and (c,d) a spatially extended initial state with
Gaussian probability distribution of width $\sigma_{HWHM}=7$, whose
pseudo-momentum is centered at one of the DPs [$\left(u_{DP},v_{DP},w_{DP}\right) =\left(\pi/4,\pi/4,-3\pi/4\right)$].
In both cases $\theta_{1}=\theta_{2}=\theta_{3}=\pi/4$ and the initial coin state is $\operatorname{col}(0,1)$.
Panel (e) shows the tripartite negativity $N^{(3)}$ against the number of time steps $t$ in  \textrm{AQW}$_{\mathrm{(3)}}$, with the walker starting at the origin and initial coin state $\operatorname{col}\left(  1,i\right)  /\sqrt{2}$.}
\label{fig2}
\end{figure}

In Figs. \ref{fig2}(a) and 2(b) we present two bidimensional projections of the probability
distribution corresponding to the propagation of a walker initially localized
at the origin. The width of the distribution grows linearly with time, along
all three spatial dimensions, as it happens for lower dimensionality. In
order to show the effect of the DP, Figs. \ref{fig2}(c) and 2(d) show the same projections as
Figs. \ref{fig2}(a) and 2(b) when the initial condition of the walker is not localized but extended. Again
we choose a Gaussian distribution with $\sigma_{HWHM}=7$, whose
pseudo-momentum is centered at one of the DPs [$\left(u_{DP}\!,\!v_{DP}\!,\!w_{DP}\right) =\left(\pi/4,\pi/4,-3\pi/4\right)$].
We observe a symmetric
ballistic dynamics in the $\left(  x_{1},x_{2}\right)  $ plane [Fig. \ref{fig2}(c)]
that resembles that of Fig. \ref{fig1}(c). Notice the existence of two concentric
bright rings, as it occurs in conical refraction \cite{BerryWilkinson}.
However, the probability is not equally symmetric in the $\left(  x_{1}
,x_{3}\right)  $ plane [Fig. \ref{fig2}(d)] which reveals an intrinsic lack of symmetry
in \textrm{AQW}$_{\mathrm{(3)}}$: indeed, it can be shown that it is not
possible to find any initial coin state that leads to a symmetric propagation
in all directions [the chosen initial coin state, $\operatorname{col}(0,1)$ for all
populated sites, leads to a symmetric distribution in the planes
with constant $x_{3}$, but not in those for which $x_{3}$ varies, as the
figures show].
One could wonder whether \textrm{AQW}$_{\mathrm{(3)}}$ is a process similar to
Grover-$\mathrm{QW}_{\mathrm{(3)}}$ as it happens for $\mathrm{N}=2$. The
answer is negative: we have derived and compared the dispersion relations of
both processes for $\mathrm{N}=3$ and they are different. Hence $\mathrm{N}=2$
is a singular case in this respect.

\section{Generation of multipartite entanglement}

Another aspect investigated in the context of two-dimensional quantum walks is
the generation and the effects of bipartite entanglement during their evolution
\cite{QWEntanglement}. Having a richer and more complex structure than the bipartite case
\cite{Horodecki}, multipartite entanglement has recently attracted a lot of
interest in the scientific community. Clearly, a feasible system able to
generate a proper amount of genuine multipartite entanglement could be a
valuable benchmark for this rapidly growing research field. Hence a most
relevant question is whether quantum walks with $\mathrm{N}>2$ do exhibit
considerable genuine multipartite entanglement or not. We have investigated this for
$\mathrm{N}=3$. We have first to trace out the state of the coin, and we are then left with a
density matrix in the composite Hilbert space $\mathcal{H}_{x_{1}}
\otimes\mathcal{H}_{x_{2}}\otimes\mathcal{H}_{x_{3}}$ (each subspace
corresponding to a direction of the walk). We evaluate the multipartite
entanglement present in this composite system by means of the tripartite
negativity \cite{Sabin}. This is defined as the geometric average of the three
negativities that are obtained by considering the three possible bipartitions
of the total system, giving $N^{(3)}=\sqrt[3]{N_{1-23}N_{2-13}N_{3-12}}$.
Here $N_{i-jk}$ is the negativity of the composite system $\left\{
i,j,k\right\}  $ corresponding to the bipartition in the subsystem $\left\{
i\right\}  $ and the subsystem $\left\{  j,k\right\}  $. Each Hilbert
direction-subspace has a dimension growing with the number of time steps, so
we use the generalization of the negativity for higher-dimensional systems (so
as to have $0\leq N\leq1$) \cite{Lee}. We have calculated $N^{(3)}$ in
\textrm{AQW}$_{\mathrm{(3)}}$, with the walker starting at the origin and
initial coin state $\operatorname{col}\left(  1,i\right)  /\sqrt{2}$, for a
number of time steps $t$ up to 10, obtaining the plot in Fig. \ref{fig2}(e). Even if
the number of time steps considered here is not so large (due to the dimension
of the total Hilbert space, the computational power required for the
evaluation grows rapidly with $t$), it is easy to check that \textrm{AQW}
$_{\mathrm{(3)}}$ is able to generate a considerable amount of genuine multipartite entanglement.
It is also interesting to notice that $N^{(3)}$ saturates rather fast.

\section{Implementability of $\mathrm{N}$-dimensional alternate quantum walk}

Let us finally discuss the implementability of \textrm{AQW}$_{\mathrm{(N)}}$.
Realizing $\mathrm{QW}_{\mathrm{(N)}}$ is quite demanding because of the
complexity of performing coin operators to transform the needed
$2^\mathrm{N}$-dimensional qudit. On the other hand, in \textrm{AQW}
$_{\mathrm{(N)}}$: (i) a single coin qubit is required independently of $N$;
(ii) two-dimensional transformations of the qubit are easy to implement
\cite{Bouwmeester,experiments,Schreiber10}; and (iii) the sequential
application of operators $\hat{D}_{j}\hat{C}_{j}$, $j=1,\ldots,N$, makes the
implementation of \textrm{AQW}$_{\mathrm{(N)}}$ similar to that of
$\mathrm{QW}_{\mathrm{(1)}}$, provided that all $\mathrm{N}$ dimensions could
be multiplexed into a single one (similarly to what Schreiber \textit{et al.}
\cite{Schreiber12} have recently done in their pioneering implementation of
$\mathrm{QW}_{\mathrm{(2)}}$). Indeed, one could even implement \textrm{AQW}
$_{\mathrm{(N)}}$ with a constant number of physical elements independently of
$\mathrm{N}$ if there is sufficient control on the experimental device.

In order to illustrate this, we generalize the idealized device already
discussed in Refs. \cite{Knight,KnightOC}, which is similar to that actually
used in Ref. \cite{Bouwmeester}. Consider a long enough optical cavity
containing two electro-optic modulators (EOMs) whose roles are (i) EOM1
performs the coin operator $\hat{C}$ (i.e., makes a unitary transformation of
the light polarization state, which plays the role of the coin-qubit in this
implementation of the walk); and (ii) EOM2 performs the conditional
displacement $\hat{D}$, which consists in up/down shifting the carrier
frequency of the light pulse depending on its polarization. The light pulses
entering the cavity are assumed to be much shorter than the cavity length, and
the frequency shifts introduced by EOM2 must be large enough for avoiding any
frequency overlapping between pulses. With such a device not only
$\mathrm{QW}_{\mathrm{(1)}}$ can be implemented -- see Refs.
\cite{Bouwmeester,Knight,KnightOC} -- but also \textrm{AQW}$_{\mathrm{(N)}}$ could
be implemented for different values of $\mathrm{N}$ by properly programming the
operations of the EOMs, without the need for additional elements. For example,
in order to perform $\textrm{AQW}_{\mathrm{(2)}}$, the first half
of the time step is implemented within one cavity round trip of the light
pulse (during which EOM1 and EOM2 implement $\hat{D}_{1}\hat{C}_{1}$); then,
in the subsequent cavity round trip, the settings of both EOMs are changed in
order to perform a different coin operator and a different frequency
displacement implementing $\hat{D}_{2}\hat{C}_{2}$. Importantly, the frequency
displacements in $\hat{C}_{1}$ and $\hat{C}_{2}$ must be different enough in
order to multiplex a large number of steps \cite{Schreiber12}. Only technical
limitations seem to restrain the extension of the procedure to higher
$\mathrm{N}$ \cite{note}. However, we are not
claiming that the device just outlined is the most appropriate for
implementing \textrm{AQW}$_{\mathrm{(N)}}$. Indeed, a suitable modification of
the flexible device used by Schreiber \textit{et al}.
\cite{Schreiber10,Schreiber12} would probably be a more promising option. With
our discussion we just want to emphasize that a single and conceptually simple
device \textit{could} implement alternate quantum walks with tunable
dimensionality.

\section{Conclusions}

In conclusion, we have introduced the $\mathrm{N}$-dimensional alternate
quantum walk and discussed some of its properties through the analysis of the
dispersion relation that reveals, in particular, the existence of diabolical
points. We have demonstrated that, for $\mathrm{N}=3$, the process
generates genuine multipartite entanglement. We have finally discussed
its implementability, that could be possible with physical resources
that do not necessarily grow with the dimensionality of the walk.

\section{Acknowledgments}

We thank Th. Busch, M. Mc Gettrick, C. M. Chandrashekar, C.
Navarrete-Benlloch, A. P\'erez, A. Romanelli, and M. Hinarejos for discussions.
This work has been supported by the Spanish Government and the European Union
FEDER through Projects No. FIS2008-06024-C03-01 and No. FIS2011-26960, by the
Basque Government through Grant No. IT472-10, and by the UK EPSRC, Grant No. EP/G004579/1 under the ``New directions for EPSRC
research leaders" initiative. C.D.F. acknowledges A. P\'erez and the
Universitat de Val\`encia for the kind hospitality.


\begin{thebibliography}{99}

\bibitem {discreteQW}Y. Aharonov, L. Davidovich, and N. Zagury, Phys. Rev. A
\textbf{48}, 1687 (1993); D. Meyer, J. Stat. Phys. \textbf{85}, 551 (1996); J.
Watrous, \textit{Proceedings of the 33rd Annual ACM Symposium on the Theory of Computing,
Heraklion, Crete, Greece, July 06-08, 2001} (ACM Press, New York,
2001), p.60.

\bibitem {continuousQW}E. Farhi and S. Gutmann, Phys. Rev. A \textbf{58}, 915
(1998); A.M. Childs and J. Goldstone, Phys. Rev. A \textbf{70}, 042312 (2004).

\bibitem {Strauch}F.W. Strauch, Phys. Rev. A \textbf{73}, 054302 (2006).

\bibitem {reviews}A. Ambainis, Int. J. Quant. Inf. \textbf{1}, 507 (2003); J.
Kempe, Contemp. Phys. \textbf{44}, 307 (2003); V. Kendon, Math. Struct. Comp.
Sci. \textbf{17}, 1169 (2006); V. Kendon, Philos. Trans. R. Soc., A \textbf{364},
3407 (2006); N. Konno, in \textit{Quantum Walks}, edited by U. Franz and M. Sch\"urmann, Quantum Potential Theory,
Lecture Notes in Mathematics Vol. 1954, (Springer, Berlin, 2008).

\bibitem {Bouwmeester}D. Bouwmeester et al., Phys. Rev. A \textbf{61}, 013410 (1999).

\bibitem {experiments}B. Do et al., J. Opt. Soc. Am. B \textbf{22}, 499 (2005);
C.A. Ryan et al., Phys. Rev. A \textbf{72}, 062317 (2005); P. Zhang et al.,
\textit{ibid} \textbf{75}, 052310 (2007); P. H. Souto Ribeiro et al.,
\textit{ibid} \textbf{78}, 012326 (2008); M. Karski et al., Science
\textbf{325}, 174 (2009); H.B. Perets et al., Phys. Rev. Lett. \textbf{100},
170506 (2008); H. Schmitz et al., \textit{ibid} \textbf{103}, 090504 (2009);
F. Z\"ahringer et al., \textit{ibid} \textbf{104}, 100503 (2010);
M.A. Broome et al., \textit{ibid} \textbf{104}, 153602
(2010).

\bibitem {Schreiber10}A. Schreiber et al., Phys. Rev. Lett. \textbf{104},
050502 (2010); A. Schreiber et al., \textit{ibid} \textbf{106}, 180403 (2011).

\bibitem {multiparticle}A. Peruzzo et al., Science \textbf{329}, 1500 (2010);
J.C.F. Matthews et al., arXiv:quant-ph/1106.1166 (2011);
J.O. Owens et al., New J. Phys. \textbf{13}, 075003 (2011); L. Sansoni et
al., Phys. Rev. Lett. \textbf{108}, 010502 (2012).

\bibitem {Mackay02}T.D. Mackay, S.D. Bartlett, L.T. Stephenson, and B.C.
Sanders, J. Phys. A: Math. Gen. \textbf{35}, 2745 (2002).

\bibitem {Moore}C. Moore and A. Russell, \textit{Proceedings of RANDOM, 2002},
edited by J.D.P. Rolim and P. Vadham (Springer, Cambridge, MA, 2002), pp.164-178.

\bibitem {Soriano}E. Rold\'an and J.C. Soriano, J. Mod. Opt. \textbf{52}, 2649 (2005).

\bibitem {Schreiber12}A. Schreiber et al., Science \textbf{336}, 55 (2012).

\bibitem {2DQW}B. Tregenna et al., New. J. Phys. \textbf{5}, 83 (2003); I.
Carneiro et al., \textit{ibid} \textbf{7}, 156 (2005); A.C. Oliveira et al.,
Phys. Rev. A \textbf{74}, 012312 (2006); K. Watabe et al., \textit{ibid}
\textbf{77}, 062331 (2008); M. Stefan\'ak et al., Phys. Scr., T \textbf{140},
014035 (2010); N. Shenvi et al., Phys. Rev. A \textbf{67}, 052307 (2003).

\bibitem {Inui}N. Inui, Y. Konishi, and N. Konno, Phys. Rev. A \textbf{69},
052323 (2004).

\bibitem {DiFranco11a}C. Di Franco, M. Mc Gettrick, and Th. Busch, Phys. Rev.
Lett. \textbf{106}, 080502 (2011).

\bibitem {DiFranco11b}C. Di Franco, M. Mc Gettrick, T. Machida, and Th. Busch,
Phys. Rev. A \textbf{84}, 042337 (2011).

\bibitem {BerryWilkinson}M.V. Berry and M. Wilkinson, Proc. R. Soc. A
\textbf{392}, 15 (1984).

\bibitem {Berry}M.V. Berry and M.R. Jeffrey, Prog. Opt. \textbf{50}, 13 (2007).

\bibitem {deValcarcel10}G.J. de Valc\'arcel, E. Rold\'an, and A. Romanelli, New J.
Phys. \textbf{12}, 123022 (2010).

\bibitem {Knight}P.L. Knight, E. Rold\'an, and J.E. Sipe, Phys. Rev. A
\textbf{68}, 020301 (2003); P.L. Knight, E. Rold\'an, and J.E. Sipe, J. Mod. Opt. \textbf{51}, 1761 (2004).

\bibitem {Marga}G.J. de Valc\'arcel, M. Hinarejos, E. Rold\'an, A. P\'erez, and A. Romanelli, arXiv:quant-ph/1212.3600 (2012).

\bibitem {spectra}C. A. Mead and D.G. Truhlar, J. Chem. Phys. \textbf{70},
2284 (1979); L.S. Cederbaum et al., Phys. Rev. Lett \textbf{90}, 013001 (2003).

\bibitem {Bostwick}A. Bostwick et al., New J. Phys. \textbf{9}, 385 (2007).

\bibitem {CastroNeto}A.H. Castro Neto et al., Rev. Mod. Phys. \textbf{81}, 109 (2009).

\bibitem {optical lattices}M. Zhang, H.H. Hung, Ch. Zhang, and C. Wu, Phys.
Rev. A \textbf{83}, 023615 (2011).

\bibitem {Young12}S.M. Young et al., Phys. Rev. Lett. \textbf{108}, 140405 (2012).

\bibitem {QWEntanglement}Y. Omar, N. Paunkovi\'c, L. Sheridan, and S. Bose, Phys. Rev. A \textbf{74}, 042304 (2006); S.E. Venegas-Andraca and S. Bose, arXiv: quant-ph/0901.3946 (2009); P. Xue and B.C. Sanders, Phys. Rev. A \textbf{85}, 022307 (2012).

\bibitem {Horodecki}R. Horodecki et al., Rev. Mod. Phys. \textbf{81}, 865 (2009).

\bibitem {Sabin}C. Sabin and G. Garcia-Alcaine, Eur. Phys. J. D \textbf{48},
435 (2008).

\bibitem {Lee}S. Lee et al., Phys. Rev. A \textbf{68}, 062304 (2003).

\bibitem {KnightOC}P.L. Knight, J.E. Sipe, and E. Rold\'an, Opt. Commun.
\textbf{227}, 147 (2003).

\bibitem {note}The multiplexing requires that the frequency displacements
corresponding to different dimensions have to be very different (in fact, this
difference determines the maximum number of steps that can be implemented; see
the discussion in Ref. \cite{Schreiber12}). Moreover, the frequency shifts should
be done with a high repetition rate (larger than the inverse of the cavity
round-trip time). All this could be very demanding for actual EOMs.

\end{thebibliography}
\end{document}